\documentclass[12pt]{elsarticle}
\journal{Current Opinion in Colloid and Interface Science }
\usepackage{amssymb}
\usepackage{amsmath}
\usepackage[symbol]{footmisc}
\usepackage[table]{xcolor}
\usepackage[autostyle]{csquotes}
\biboptions{sort&compress}

\begin{document}

\begin{frontmatter}

%% Title, authors, and addresses
\title{Simulating dynamic bonding in soft materials}

%% Author name
\author[a]{Tyla R. Holoman} 
\author [b] {B. P. Prajwal}
\author[d,e]{Glen M. Hocky}
\author[a,b,c]{Thomas M. Truskett}

%% Author affiliation
\affiliation[a]{organization={McKetta Department of Chemical Engineering, University of Texas at Austin},
             city={Austin},
             state={Texas},
             postcode={78712},
             country={United States}}

\affiliation[b]{organization={Department of Chemical Engineering, University of Michigan},%Department and Organization
            city={Ann Arbor}, 
            state={Michigan},
            postcode={48109},
            country={United States}}

\affiliation[c]{organization={Biointerfaces Institute, University of Michigan},%Department and Organization
            city={Ann Arbor}, 
            state={Michigan},
            postcode={48109},
            country={United States}}

\affiliation[d]{organization={Department of Chemistry, New York University},%Department and Organization
            city={New York City},
            state={New York},
             postcode={10003}, 
            country={United States}}

\affiliation[e]{organization={Simons Center For Computational Physical Chemistry, New York University},%Department and Organization 
            city={New York City},
            state={New York},
             postcode={10003}, 
            country={United States}}

%% Abstract
\begin{abstract}
Dynamic bonding is an essential feature of many soft materials. Molecular simulations have proven to be a powerful tool for modeling bonding kinetics and thermodynamics in these materials, providing insights into their properties that cannot be obtained by experiments alone. Here, we review recent advances in modeling dynamic bonding in soft matter via molecular dynamics, Monte Carlo, and hybrid simulation methods, highlighting outstanding challenges and future directions. 
\end{abstract}

%% Keywords
\begin{keyword}
molecular dynamics \sep Monte Carlo \sep dynamic bond exchange \sep covalent adaptable networks \sep cytoskeletal assembly
\end{keyword}

\end{frontmatter}

\section*{Introduction}
\label{intro}

\begin{figure}
    \centering
    \includegraphics[width=0.9\linewidth]{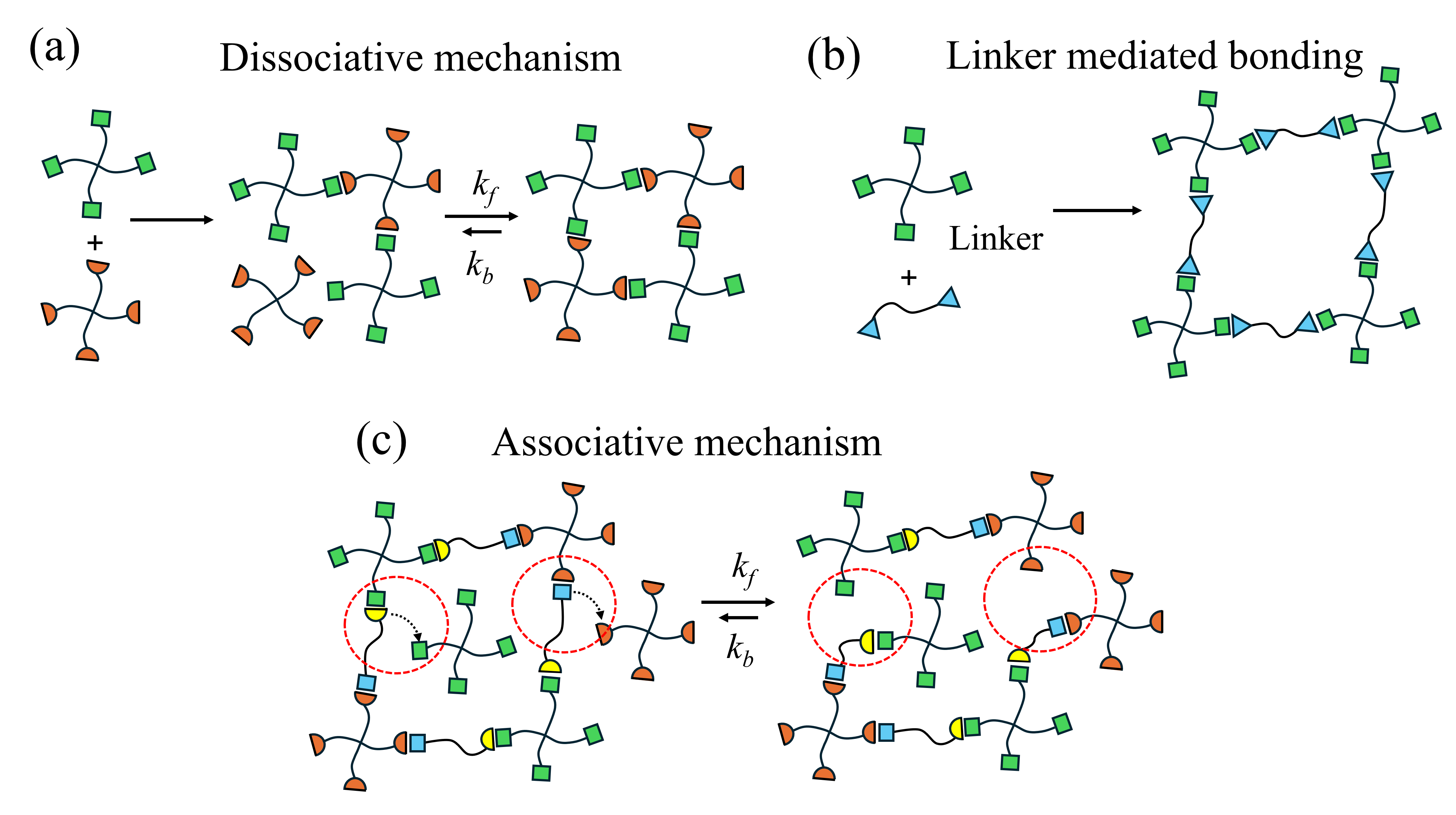}
    \caption{Mechanisms associated with network formation and reversible rearrangement using star polymers as building blocks. a) Reversible dissociative bonding process between two types of star polymers, A (green) and B (red). b) Linker-mediated bonding of star polymers. c) Reversible associative bonding for a linker-mediated star polymer network.}
    \label{fig:bond_schematic}
\end{figure}

Many soft materials comprise molecular or colloidal building blocks that assemble through reversible supramolecular or dynamic covalent bonds, resulting in superstructures such as clusters, bundles, and percolating gel networks. These superstructures exhibit emergent behaviors that reflect the complex interplay between the properties of the building blocks and the thermodynamics and kinetics of the bonding interactions. The building blocks in these systems encompass a diverse range of components, including synthetic colloids, polymers, biomolecules (such as actin, tubulin, or streptavidin), and small molecules. The reversible dynamic bonds that form between these building blocks can be classified as dissociative or associative, depending on the mechanism for bond rearrangement in the material (Fig.~\ref{fig:bond_schematic}). Dissociative bonds are characterized by formation, breakage, and reformation processes between the same or different bonding pairs. As a result, the number of bonds changes over time and can be further modulated in the presence of a stimulus that shifts the bonding equilibrium, such as changes in temperature or pH. In contrast, associative bonding involves swapping between neighboring bonding partners while conserving the total number of bonds during the exchange. This mechanism enables continuous structural rearrangements within networks linked by associative bonds; however, the changes are topologically constrained by network connectivity, and the rate of change is limited by the kinetics of associative swaps. Both dissociative and associative mechanisms form covalent adaptable networks (CANs) with building blocks linked by dynamic covalent bonds~\cite{Kloxin2013CovalentSystems, Webber2022DynamicInteractions}. Polymeric associative networks are commonly referred to as vitrimers, and modeling approaches that explain how dynamic bond exchanges influence their macroscopic properties have been recently reviewed elsewhere~\cite{Karatrantos2024MolecularReview}. 

Reversible dynamic bonding enables precise tuning of the structure and corresponding physical properties of soft materials~\cite{Zhang2022Structure-Reactivity-PropertyNetworks, dominguez2020assembly,kang2022colorimetric,kang2025colloidal,kim2025dense}, thus promoting functionalities such as self-healing, responsiveness, reprocessability, and recyclability~\cite{Bertsch2023Self-HealingRegeneration,wang2020self}. In synthetic polymer networks, for example, dynamic crosslinking can be systematically modulated by molecular substitution of the bonding pair, changing pH, or using linker mixtures to control the mechanical properties and viscoelastic response~\cite{FitzSimons2020PreferentialAdditions, Yesilyurt2017MixedNetworks, crowell2023shear, tang2021dynamic, xu2011divergent,crolais2023enhancing}. Similarly, biopolymer networks, such as the actin cytoskeleton, exhibit sequence-defined binding characteristics, giving rise to adaptive viscoelastic behavior that is crucial for cellular processes~\cite{wachsstock1994cross,lieleg2010structure}. Overall, by leveraging reversible dynamic bonding at the molecular scale, mesoscale structures with functional properties can be designed in synthetic polymeric and biological materials for potential applications in 3D printing, tissue engineering, and drug delivery~\cite{rosales2016design,loebel2017shear}. 

Various analytical experimental techniques are available to characterize dynamic and reversible bonding in soft materials~\cite{Brandt2014State-of-the-ArtMaterials}. Nonetheless, a strong scientific basis for the microscopic interpretation of experiments, including an understanding of topological defects, requires complementary molecular modeling. Computational and theoretical methods to address this challenge are under development and have made significant strides, particularly for crosslinked polymer networks. To this end, for example, Monte Carlo simulations have been used in conjunction with proton multiple-quantum nuclear magnetic resonance (NMR) measurements at low fields to characterize the bond networks formed by end-functionalized star polymers~\cite{Lange2011ConnectivityStudy}. Similarly, polymer network theories have been developed together with computer simulations to gain insights on topological defects, elastic response, and gelation of star-polymer networks~\cite{Zhong2016QuantifyingElasticity, Lin2019RevisitingNetworks}.

\begin{figure}
    \centering
    \includegraphics[width=0.8\linewidth]{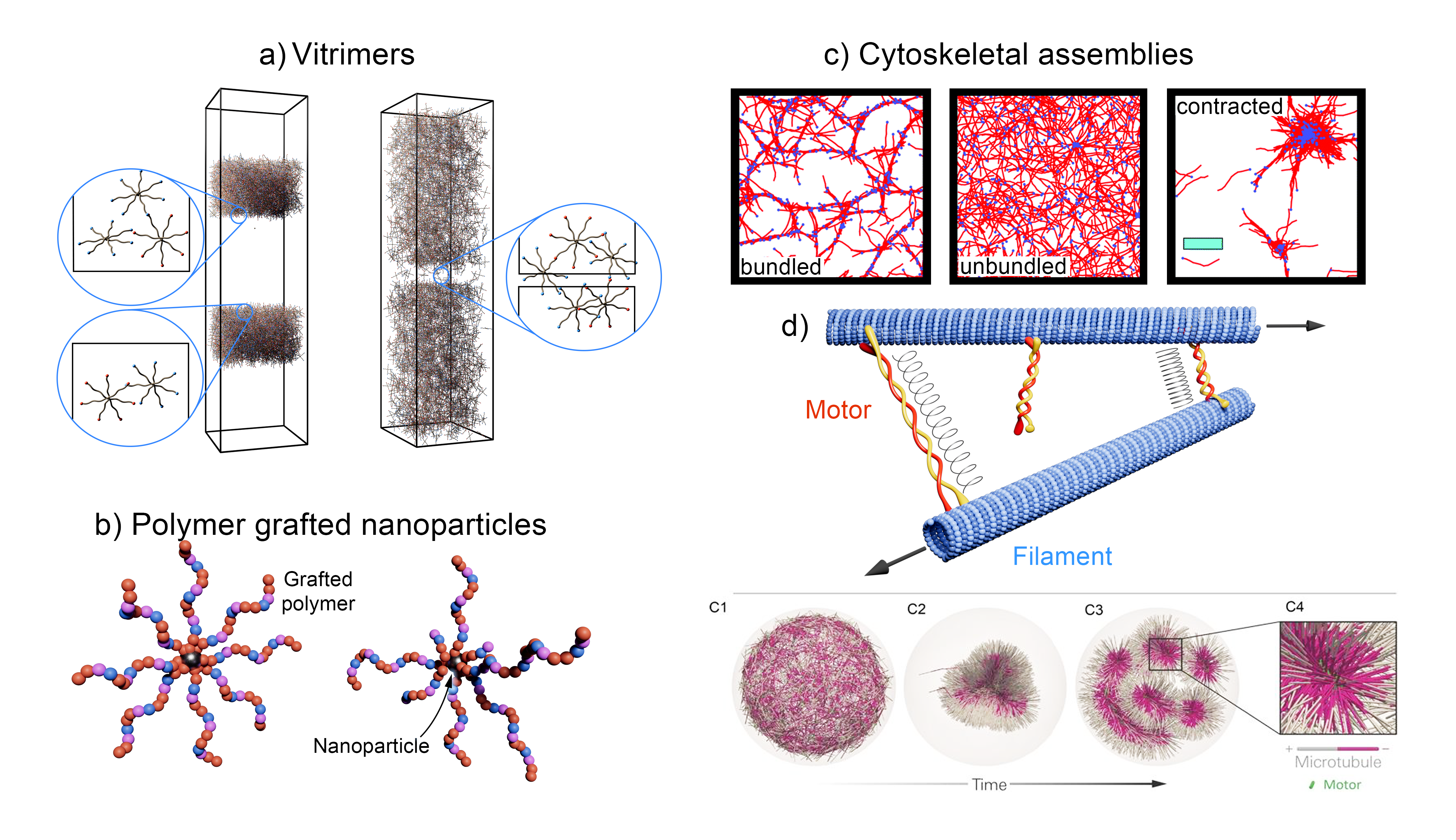}
    \caption{Examples of simulated dynamic soft matter. a) Self-healing vitrimers (reproduced from~\cite{Ciarella2019Swap-DrivenVitrimers}). b) Polymer-grafted nanoparticles with bond exchange reactions (reproduced from~\cite{Chen2024TopologicalExchanging}). c) Actomyosin networks, simulated with AFINES (adapted from \cite{Freedman2018NonequilibriumNetworks}). d) Cytoskeletal assembly, modeled using kinetic Monte Carlo simulations (reproduced from~\cite{Yan2022TowardAssemblies}).}
    \label{fig:summary}
\end{figure}

Recent efforts have focused on developing a comprehensive simulation framework capable of predicting the impact of specific, reversible, bonding interactions on the structure and emergent properties in a diverse variety of soft materials (Fig.~\ref{fig:summary}). A key challenge lies in accurately incorporating bonding kinetics, which emerges from molecular-scale processes, but impacts mesoscale structural evolution of the bonded networks.
The time scales for specific bonding interactions may be comparable or significantly different than those associated with the formation or reconfiguration of the network. Furthermore, the concentration influences the local organization of the molecular or colloidal building blocks, which in turn affects the kinetics of bond formation and the corresponding reaction rates. These conflicting demands on spatiotemporal resolution are primarily addressed through simulations from three broad categories: molecular dynamics (MD), Monte Carlo (MC), and hybrid approaches that combine MD and MC. Within each of these categories, it is possible to simulate associative or dissociative interactions with tunable kinetics using coarse-grained models with varying levels of resolution. Although many foundational modeling methods and theories have been developed over the past few decades, substantial progress has been made recently in enhancing and expanding these models to interpret experimental data obtained from materials with properties relevant to multiple application spaces. 

\section*{Molecular dynamics}
\label{MD}

\subsection*{Coarse-grained models}
\label{md_cg}
Various methods exist for modeling controllable reversible bonding in coarse-grained molecular dynamics simulations. Coarse-grained models achieve computational efficiency by removing atomistic detail, allowing for larger time-step simulations that capture ``rare'' bonding events and structural rearrangements. These models also facilitate the simulation of large system sizes, elucidating the multiscale structural evolution of network formation. 

The simplest coarse-grained model to simulate dynamic reversible bonding involves using a bonding site with an isotropic interaction potential with a relatively shallow attractive well. This method has been widely used within the soft-matter community and implemented in open-source modeling packages. Ensuring specific, monofunctional bonding, while possible (see, e.g., \cite{bianchi2006phase, howard2019structure}), is challenging to enforce with such nonspecific potentials, as is the independent parameterization of bonding thermodynamics and kinetics. These shortcomings limit their use for studying assembly or structural evolution processes that rely on bond swapping, breaking, and formation. In this section, we examine MD methods for addressing these shortcomings for specific classes of bonding phenomena in soft materials.   

\subsubsection*{RevCross three-body potential}
\label{3body}

The RevCross three-body potential has been widely used to simulate vitrimers in MD simulations with recent focus on bond-exchange dynamics, as well as mechanical and self-healing properties~\cite{Ciarella2019Swap-DrivenVitrimers, Ciarella2022AssociativeDynamics, Li2024ManipulatingBonds, lin2025molecular, zhao2022molecular}. The RevCross potential has also been used to understand how entropy can drive inter-polymer versus intra-polymer reversible bond formation between chains with spacer-sticker sequences. These simulations clarify the entropic origin of attractions between self-complementary DNA-grafted nanoparticles~\cite{sciortino2020combinatorial} and single-chain nanoparticles~\cite{rovigatti2022designing}, as well as phase transitions in soft and biological polymer networks~\cite{rovigatti2023entropy, tosti2025entropy}. 

One of the advantages of using the RevCross three-body potential is that the bond swapping rate can be modulated by a single parameter, which serves as a means to tune the mechanical responses of the material. In this novel method, associative bond-swap processes are modeled in a continuous MD framework through a repulsive three-body potential~\cite{Sciortino2017Three-bodyDynamics},
\begin{equation}
    V_\text{three-body} = \lambda \sum_{ijk}\epsilon V_3(r_{ij})V_3(r_{ik})
\end{equation}
where $i$ is the central particle, and $r_{ij}$ and $r_{ik}$ are the pair-wise distances between bonding candidates $j$ and $k$, respectively. This potential is added to any short-range pairwise bond potential $V_\text{bond}(r)$ with minimum energy $V_\text{bond} (r_{\text{min}}) = -\epsilon$. The parameter $\lambda$ creates a swap energy barrier $\epsilon (\lambda-1)$, and $V_3 (r)$ is given by
\begin{equation}
    V_3(r) = \begin{cases}
        1, & r \leq r_\text{min} \\
        -{V_\text{bond}(r)}/{\epsilon}, & r_\text{min} \leq r \leq r_\text{cutoff}
    \end{cases}
\end{equation}
 When $\lambda = 1$, bonds can swap spontaneously, and for $\lambda \gg 1$, the swap barrier increases, making it more difficult for particles to swap bonds in the system. This term $V_3 (r)$ ensures a one-to-one bonding scheme as long as $\lambda \geq 1$. If $\lambda < 1$, the three-body term becomes too small to counteract the attractive forces, leading to non-physical clustering in the system~\cite{Ciarella2022AssociativeDynamics}. An advantage of using the three-body potential is that the particles can swap ``virtual'' bonds without causing a significant change in the total potential energy of the system, revealing entropic effects. The three-body associative bonding potential is implemented in HOOMD-Blue, an open-source simulation software well-known for its compatibility with GPUs~\cite{Ciarella2022AssociativeDynamics, Anderson2020HOOMD-blue:Simulations}. The package can be found in HOOMD-Blue's MD module as a many-body potential titled RevCross. 

\subsubsection*{REACTER framework}
\label{reacter}

To incorporate more details into modeling the dynamic bonding mechanism, a method coined as REACTER was developed in the Large-scale Atomic/Molecular Massively Parallel Simulator (LAMMPS) open-source package ~\cite{Gissinger2017ModelingSimulations, Thompson2022LAMMPS}. REACTER is a versatile approach that has been frequently used to simulate coarse-grained vitrimer models~\cite{Zhao2022MolecularNetwork, Chen2024TopologicalExchanging, Singh2024UnderstandingSimulations, Singh2020ModelingCapability}. A key benefit of REACTER's reaction template method is its flexibility in allowing user-defined dynamic reactions with various potentials, including atomistic force fields. 

REACTER leverages pre- and post-reaction templates to integrate chemical reactions during MD simulations on the fly. Compared to other methods, the REACTER algorithm is distinguished by its description of reactions as changes to the topologies of reaction sites. This approach requires the user to specify the pre- and post-reaction topologies in template files along with a reaction map file that describes how atoms map between the templated states. Within REACTER, the superimpose algorithm is responsible for locating pre-reaction topologies in the system. The superimpose algorithm begins when two bonding atom types are separated by less than a specified bonding cutoff distance. These two atoms are then matched to their locations in the pre-reaction template, and the pre-reaction topology is confirmed. Upon finding the pre-reaction topology, the reaction changes the topology of the selected particles to the post-reaction topology, where the standard dynamics are performed on the new topology. Initially, REACTER lacked parallelization and relied solely on a probability-based approach to model reaction events, which was incapable of modeling bond breaking. REACTER's capabilities were expanded in 2020 to include parallelization, bond-breaking reactions, and the ability to model additional reaction constraints, such as angular or dihedral restrictions, providing users with more control over the modeled reaction~\cite{Gissinger2020Reacter:Dynamics}. These extra features, together with other capabilities added more recently~\cite{Gissinger2024MolecularREACTER}, have made REACTER a suitable candidate for modeling both associative and dissociative bonding schemes across a wide range of materials and potential applications.

\subsection*{Atomistic force fields}
\label{atomistic}
Coarse-grained MD simulations with dynamic bonding provide valuable insights into understanding how the reaction kinetics and thermodynamics relate to network formation and also provide mechanistic insights into nano- to mesoscale structural rearrangements in soft materials. However, they are limited in their ability to capture higher-resolution features of reaction pathways or chemical interactions during assembly. Atomistic simulations offer an opportunity to gain a deeper understanding of these aspects, but at a significant computational cost.

Atomistic details can be incorporated by integrating quantum mechanics (QM) calculations into classical molecular mechanics (MM) simulations of reacting systems~\cite{warshel1976theoretical}. In hybrid QM/MM approaches, small subsystems requiring detailed quantum mechanical treatment are embedded within a larger classical environment, where a much more computationally efficient force-field treatment suffices. The quality of QM/MM simulation predictions for molecular dynamics depends on many factors, including the subsystem definition, the level of QM calculation employed, the force fields adopted, and the quality of the conformational sampling achieved~\cite{brunk2015mixed,clemente2023best,ho2024accurate}. The successful application of QM/MM methods to model condensed-phase biochemical and catalytic processes at the nanoscale has inspired recent development of approaches capable of addressing similar challenges in network-forming soft materials, where the challenges related to spatiotemporal resolution are even more formidable~\cite{wang2023bringing}.

One alternative to QM/MM methods to model materials with dynamic bonding is to employ a reactive atomistic force field, such as ReaxFF, which can allow us to capture configurational rearrangements involving significantly longer length and time scales. ReaxFF was initially developed to model hydrocarbon combustion in 2001~\cite{vanDuin2001ReaxFF:Hydrocarbons} and has since undergone many updates to improve its ability to predict complex reaction pathways in agreement with QM prediction and chemical intuition~\cite{Chenoweth2008ReaxFFOxidation}, enhance accuracy, and reduce computational cost~\cite{leven2021recent}.   

ReaxFF represents the reactive interactions between atoms by using interatomic distances to calculate bond order~\cite{Senftle2016TheDirections}. An example ReaxFF potential comprises a sum of energy contributions listed below:
\begin{equation}
    E_\text{system} = E_\text{bond} + E_\text{over} + E_\text{under} + E_\text{lp} + E_\text{val} + E_\text{tors} + E_\text{vdW} + E_\text{Coulomb} 
\end{equation}
Here, $E_\text{bond}$ is the bond energy, $E_\text{over}$ and $E_\text{under}$ represent over- and under-coordination penalties, respectively. $E_\text{lp}$ is the lone-pair energy, $E_\text{tors}$ and $E_\text{val}$ are the torsion- and valence-angle energies, respectively, and $E_\text{vdW}$ and $E_\text{Coulomb}$ represent the dispersive and electrostatic interactions between all atoms. Depending on the system, terms representing other energetic contributions, e.g., hydrogen bonding, are also included. 
ReaxFF can be used to simulate a diverse array of reactive materials and has been implemented in simulation software packages, including LAMMPS~\cite{Thompson2022LAMMPS}. A challenge lies in modeling many polymer cross-linking reactions since they occur on timescales of minutes to hours, which is not feasible to access in simulations using atomistic models. To simulate such chemical reactions with slower kinetics, Vashisth and coworkers~\cite{Vashisth2018AcceleratedPolymers} introduced an accelerated ReaxFF method, which was used to model polymeric materials undergoing cross-linking reactions with higher energetic barriers. 

Accelerated ReaxFF overcomes this barrier by providing a ``bond-boost'' to the reactive atoms when they come within a specified distance from each other~\cite{Miron2003AcceleratedMethod}. The bond boost or so-called restraint energy ($E_\text{rest}$) is effectively another energetic contribution that either stretches or compresses the bond distance between the atoms, enabling the bond formation or breaking process between the atoms. 
$E_\text{rest}$ is dependent on the particles' interatomic distance, $r_{12}$ and two force parameters, $F_1$ and $F_2$, and is given as, 
\begin{equation}
    E_\text{rest} = F_1 \{ 1-\text{exp}[{-F_2(r_{ij}-r_{12})^2}] \}
\end{equation}
where $r_{ij}$ is the actual distance between particles~\cite{Vashisth2018AcceleratedPolymers}. This boost is implemented in a way that allows for cross-linking to be readily simulated under realistic, low-temperature conditions, producing results consistent with experimental data with energy barriers similar to those computed from QM. Recent studies have used the accelerated ReaxFF method to model crosslinked epoxide-amine polymer networks~\cite{Vashisth2018AcceleratedPolymers}, vitrimers~\cite{Zheng2025AcceleratedNetworks}, and CANs with Diels-Alder reactions~\cite{Vermeersch2024ComputationalRecyclability}. 

\section*{Monte Carlo}
\label{mc}
\begin{figure}
    \centering
    \includegraphics[width=1\linewidth]{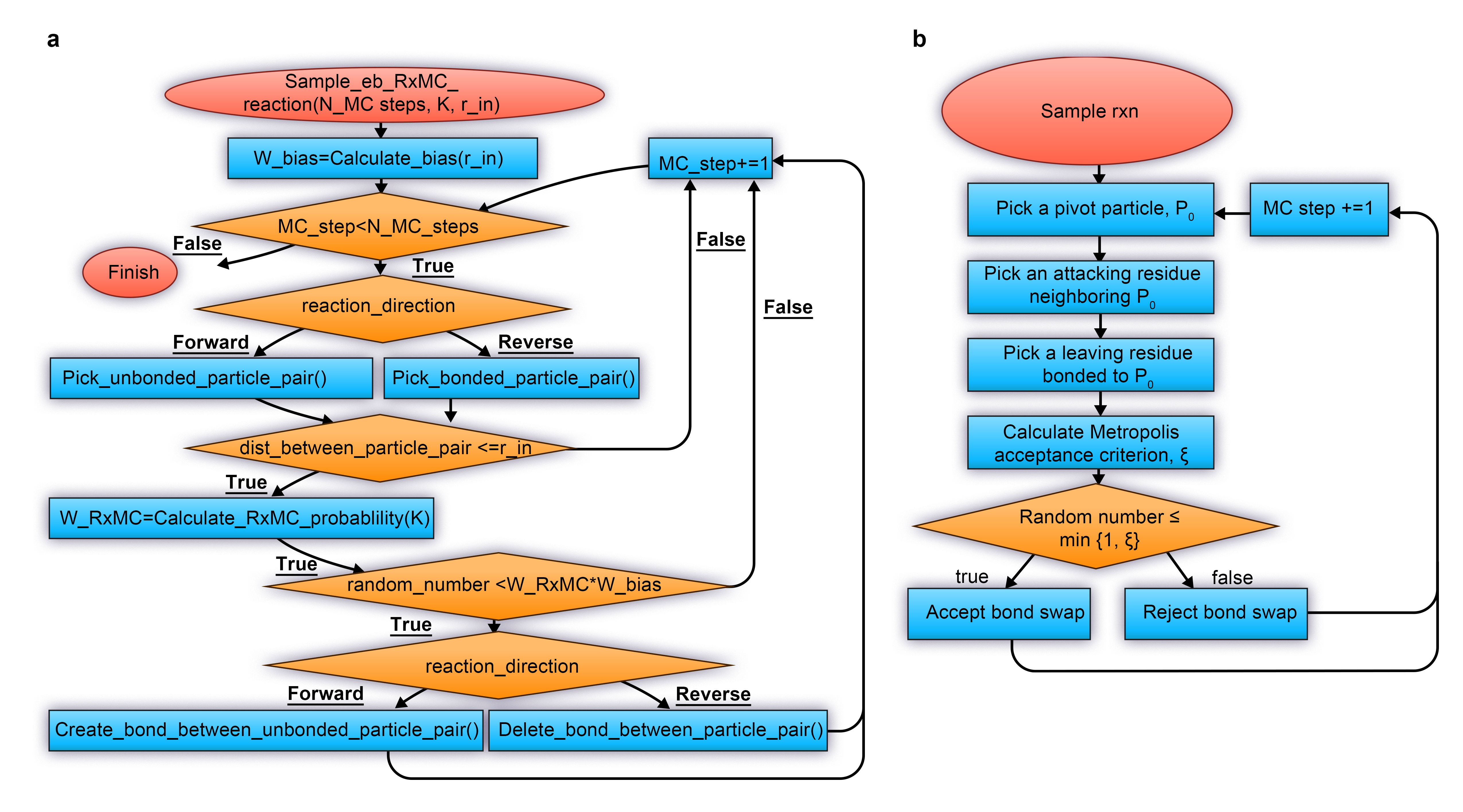}
    \caption{Comparison of Monte Carlo algorithms for modeling a) dissociative (reproduced from~\cite{Blanco2024TheMethod}) and b) associative (reproduced from~\cite{Rao2024APolymers}) interactions.}
    \label{fig:MC_algos}
\end{figure}

Monte Carlo methods have been widely employed to model reversible, dynamic bonding reactions in soft materials. The formation of bonds can be modeled using Reactive Canonical Monte Carlo (RCMC). This method selects potential molecules to react and then accepts reaction moves to form products with a probability that accounts for the ideal-gas energetic differences between isolated products and reactants, as well as the interactions these species have with their surroundings. The latter may include interactions with solvent, other reactant or product molecules, or external fields~\cite{Johnson1994ReactiveCarlo, Smith1994TheExamples}. RCMC has been used extensively to study chemical reactions in complex fluids~\cite{Turner2008SimulationReview}. Recently, RCMC has been extended to the explicit bonding reaction ensemble Monte Carlo (eb-RxMC) method, which adds and removes bonds explicitly rather than inserting or removing molecules and complexes~\cite{Blanco2024TheMethod}. eb-RxMC is currently implemented in ESPReSso, and the eb-RxMC algorithm for modeling dissociative bonding is shown in Fig.~\ref{fig:MC_algos}a. 

Monte Carlo simulations can also be used to model associative bond-swapping interactions. Figure~\ref{fig:MC_algos}b shows an example algorithm, adapted from Ref.~\citenum{Rao2024APolymers}. This approach is more versatile compared to previous associative modeling approaches, such as Ref.~\citenum{Sciortino2017Three-bodyDynamics}, due to its ability to treat multivalent, multi-species interactions. Although the three-body bond-swapping potential is a convenient choice for modeling the dynamics of monovalent associative bond exchanges in dilute systems, it cannot handle multivalent bonding schemes and does not accurately capture the physics of reactive systems at high densities, where collisions and particle swaps happen incessantly. 
On the other hand, in the Monte Carlo algorithm developed by Rao and coworkers~\cite{Rao2024APolymers}, detailed balance is preserved for multivalent interactions by introducing a bias term to the MC move so that its acceptance probability is given as,
\begin{equation}
    P_\mathrm{acc} = \text{min}\left[1, \frac{p'/v'^u}{p/v^u}\exp(-\beta \Delta G_0)\right] 
\end{equation}
Here, $p$ and $p'$ are the forward and reverse proposal probabilities for a bond swap pair, $v^u$ and $v'^u$ are the unoccupied valences for the attacking and leaving residues (see Fig.~\ref{fig:MC_algos}A), $\beta = 1/k_\text{B}T$, $k_\text{B}$ is the Boltzmann constant, $T$ is temperature, and $\Delta G_0$ is the change in reaction free energy~\cite{Rao2024APolymers, Xia2022Entropy-DrivenVitrimer}. 

\section*{Hybrid MD/MC}
\label{mdmc}

Although the MC algorithms described above are designed to sample equilibrium configurations, MC can also be combined with MD simulations that advance configuration trajectories between reaction moves, forming hybrid MD/MC methods.
Such hybrid algorithms have been employed to explore the dynamic evolution of CANs and other soft materials linked by reversible bonds.

One of the foundational implementations of MD/MC bonding methods in the polymer physics community is the algorithm of Hoy and Fredrickson~\cite{Hoy2009ThermoreversiblePhysics}. The method introduced by these authors was designed to simulate thermoreversible networks of coarse-grained (i.e., Kremer-Grest) polymers with a fraction of ``sticky'' monomers that can be crosslinked by reversible supramolecular bonds. In this method, particle trajectories are advanced using Langevin dynamics, and every $n$ timesteps, the MD integration is paused to perform the MC portion of the algorithm. An equal fraction of bonded and unbonded sticky monomer pairs are then chosen for bonding and unbonding moves; this fraction allows kinetics to be tuned independent of bond thermodynamics. Sticky bonds are modeled by the potential 
\begin{equation}
    U_\text{sb}(r,h) = U_\text{FENE}(r)-U_\text{FENE}(r_0)-h
\end{equation}
which utilizes the finite extensible nonlinear elastic (FENE) bond potential and includes the equilibrium bond length $r_0 \approx 0.96\sigma$, where $\sigma$ is the bead diameter. The parameter $h$ represents the binding energy. For $h=0$, sticky bonds can form spontaneously, and $h$ can be tuned to control the reaction thermodynamics. 
 
Many studies have used the Hoy-Fredrickson MD/MC framework to understand various phenomena, such as epoxy curing, and the framework is implemented as a plugin to the HOOMD-Blue simulation package~\cite{Thomas2018RoutineDynamics}. The algorithm was subsequently adapted for modeling reversible dissociative bonding in more general linked colloidal systems, with an updated treatment of reaction kinetics~\cite{Mitra2023ABinders} and a new HOOMD-Blue plugin called dybond. To tune the barrier of reaction, dybond defines $P_\text{on}$ and $P_\text{off}$, the probability of forward and reverse reactions, respectively. These probabilities depend on user-defined kinetic rate constants $k_\text{on}$ and $k_\text{off}$, enabling independent tuning of the forward and reverse reactions, respectively. Dybond was initially developed to simulate binding between DNA-coated emulsion droplets~\cite{Mitra2023ABinders}, an application that highlights its capabilities for studying kinetically controlled assembly of soft materials. A similar approach was developed in parallel and has been implemented using Theoretically Informed Langevin Dynamics (TILD) simulations in the open-source simulation package MATILDA.FT to understand the phase behavior of complex coacervates~\cite{Jedlinska2024EffectsCoacervates, Jedlinska2023MATILDA.FT:Matter}.

Although Hoy \cite{Hoy2009ThermoreversiblePhysics} and many other authors (see, e.g., \cite{li2022distribution,karmakar2025computer}) have subsequently used hybrid MD/MC methods to model dissociative interactions, MD/MC methods can also be used to model associative interactions, by swapping bonds during the MC moves rather than forming and breaking them~\cite{Perego2020VolumetricStudy, Perego2022MicroscopicVitrimers, Stukalin2013Self-HealingBonds}.

\subsection*{Kinetic Monte Carlo}
\label{kMC}
While Kinetic Monte Carlo (kMC) methods have been leveraged to simulate synthetic polymer networks, such as vitrimers~\cite{Liu2024TheApproach}, they have primarily been used to model soft, biological materials with reversible networks of proteins or oligonucleotides. While these building blocks are themselves covalently linked polymers, their properties and interactions with other molecules are dictated by non-covalent interactions (especially Coulombic forces, dispersion, hydrogen bonding, and the effective hydrophobic interaction). Because these interactions are non-covalent, close associations that can be considered single `bonds' (such as a hybridized pair of DNA strands) can break and form at room temperature. 

Soft materials formed from biomolecules have functional mechanical properties that derive from their crosslinked networks. In most simulation studies, polymers are assumed to be pre-formed, and they are linked to one another by crosslinking molecules that are either explicitly included in the simulation or appear and disappear between biopolymers separated by the correct distance and exhibiting a mutually favorable orientation. 
In these systems, reversible bonds are often treated through the explicit forming and breaking of bonds (usually harmonic), thereby mixing MD and MC in another type of hybrid simulation sometimes called an ``agent-based model''. 
An interesting feature of these systems, not generally present in synthetic polymer systems, is the presence of non-equilibrium activity, either through the polymerization and depolymerization of the polymers themselves or the role of `active' crosslinkers. Active crosslinkers serve as substitutes for motor proteins, which move unidirectionally along cytoskeletal filaments. Their presence is typically modeled by a shifting location where the explicit harmonic bond is attached along the polymer at a fixed or force-dependent rate.

Actin is a biopolymer that forms fiber networks cross-linked with proteins, which can be either passive (cross-linkers) or active (motors). An example of an active crosslink, or motor, in actin networks is myosin. This protein can utilize energy from ATP hydrolysis to ``walk'' along actin filaments, a phenomenon that leads to muscle contraction. Actin is a well-understood system, making it ideal for developing new simulation methods that can encompass the complexities of similar, less-understood cytoskeletal assemblies. Several simulation methods have been developed to model systems like actin. One example is Active Filament Network Simulation (AFINES), an open-source C++ package that utilizes kinetic Monte Carlo to simulate biopolymers, crosslinkers, and motors explicitly~\cite{Freedman2017ANetworks}. With AFINES, the actin filaments are modeled as polar worm-like chains (WLC), and the crosslinkers and motors are modeled as Hookean springs with ends that can bind and unbind to the actin filaments, with the motors also having the ability to walk along the filaments. Crosslink bonding in this method is modeled using MC: at each timestep, all available filament links are cataloged, and for each link, the Metropolis factor is calculated to move the crosslinker to the nearest available filament link. The Metropolis factor is given as,
\begin{equation}
    P^{\text{off}\rightarrow\text{on}}_{xl,i} = \min [ 1, \exp(-\Delta U^\text{stretch}_{xl,i}/k_\text{B}T) ]
\end{equation}
and the bond forms with probability $(k^\text{on}_{xl}\Delta t)P^{\text{off}\rightarrow\text{on}}_{xl,i}$ or stays unbound with probability $1-\sum_i(k^\text{on}_{xl}\Delta t)P^{\text{off}\rightarrow\text{on}}_{xl,i}$. Motor proteins bind and unbind to the filaments the same way, but are also given a positive velocity, allowing them to walk along the filament,
\begin{equation}
    v(\overrightarrow{F}_m) = v_0 \max \{ 1+ {\overrightarrow{F}_m \cdot \hat{r}}/{F_s}, 0\} 
\end{equation}
where $v_0$ is the unloaded motor speed, $F_m$ is the spring force on the motor, and $\hat{r}$ is the tangent to the filament. AFINES has enabled the computation of nonequilibrium phase diagrams in actomyosin networks~\cite{Freedman2018NonequilibriumNetworks}.

There are many other similar methods, including Cytosim, an open-source simulation suite developed initially to simulate microtubules that has since been adapted for actin networks~\cite{Lugo2023ASystems}. Cytosim is a numerical simulation method that utilizes Brownian dynamics to simulate filaments and their associated proteins in one, two, or three dimensions. Another similar method is the A Living Ensemble Simulator (aLENS) package~\cite{Yan2022TowardAssemblies, Najma2025ArrestedMotors}, which differs from other methods in its handling of steric interactions between filaments in the system. While most methods rely on a purely repulsive potential between the filaments, aLENS instead uses a constraint method to prevent filaments from overlapping~\cite{Yan2019ComputingMethod}. Enforcing the hard-core repulsion with steric constraints rather than a repulsive potential allows for a larger time step and therefore longer simulations.  In contrast to packages like Cytosim or AFINES, aLENS can also include the effects of hydrodynamics.

An alternative method for simulating biopolymer assembly with crosslinks and motors is the software package Mechanochemical Dynamics of Active Networks (MEDYAN)~\cite{Popov2016MEDYAN:Networks}. MEDYAN can be used to study the mechanochemical properties of biopolymer networks by utilizing a modified Gillespie algorithm. MEDYAN differs from other simulation methods because it accounts for changing numbers of components based on the reaction dynamics. 

\section*{Modeling challenges and outlook}
\label{challenge+future}
Despite recent advancements in modeling capabilities for reversibly bonded soft materials discussed here (Table~\ref{tab:table}), there are still outstanding challenge areas ripe for further research. For example, when simulating dynamic processes using MC or hybrid MC/MD methods, moves should be chosen such that they leave the equilibrium Boltzmann distribution invariant. This criterion will be met by MC moves that satisfy detailed balance, but also by moves that satisfy a weaker balance condition consistent with regular, ergodic sampling~\cite{Manousiouthakis1999StrictSimulation}. Although most existing bond swap algorithms for monovalent particles satisfy such balance conditions, choosing appropriate MC moves for bond breaking and formation in multivalent or multi-species materials, which are common in soft matter, is less trivial~\cite{Rao2024APolymers}. Further complexity is introduced if one proposes multiple bond changes within a single MC step, in which case it is necessary to ensure that the proposed moves are independent or that the coupled probabilities are taken into account. Such scenarios, therefore, require careful consideration of forward and backward proposal probabilities to maintain balance. In contrast to bond swapping, dissociative mechanisms pose an additional challenge due to the difference in probabilities for finding unbonded versus bonded particle pairs within the reaction distance. To address this, an additional bias term (see Fig.~\ref{fig:MC_algos}a) can be included in the Metropolis acceptance criteria to ensure detailed balance is satisfied. Along with system complexity, computational efficiency should also be considered, 
as MC moves are typically performed on a single processor. Thus, although MC and hybrid MD/MC methods offer considerable flexibility for simulating dynamic bonding systems, a need remains for general strategies to determine appropriate and computationally efficient MC moves for simulating multicomponent networks with competitive bonding~\cite{hatch2025best}.

\begin{table}[t]
    \caption{Summary of simulation methods to model dynamic reversible bonding.}
    \centering
    \rowcolors{2}{lightgray!30}{white}
    \setlength{\arrayrulewidth}{0.3mm}
    \begin{tabular}{p{0.18\linewidth} p{0.12\linewidth} p{0.18\linewidth} p{0.4\linewidth} p{0.05\linewidth}}
    \rowcolor{lightgray!70}
        \textbf{Algorithm} & \textbf{Method} & \textbf{Applications} & \textbf{Implementation} & \textbf{Ref.} \\
        \hline
        RevCross$^{\dagger}$ & MD & vitrimers & included in HOOMD-Blue distribution & \cite{Ciarella2022AssociativeDynamics} \\ 
        REACTER$^{\dagger, \ddagger}$ & MD & vitrimers, polymers & included in LAMMPS distribution & \cite{Gissinger2017ModelingSimulations} \\ 
        Accelerated ReaxFF$^{\dagger, \ddagger}$ & MD & polymers, vitrimers & modified LAMMPS plug-in & \cite{Vashisth2018AcceleratedPolymers}\\
        eb-RxMC$^{\ddagger}$ & MC & polymers & Integrated with langevin dynamics in ESPReSso & \cite{Blanco2024TheMethod} \\
        Multivalent bond swaps\textsuperscript{\textdagger} & MC & vitrimers  & Potential for hybrid MD/MC implementation& \cite{Rao2024APolymers} \\
        dybond$^{\ddagger}$ & MD/MC & polymers, colloids & HOOMD-Blue plug-in available on GitHub & \cite{Mitra2023ABinders} \\ 
        TILD$^{\ddagger}$ & MD/MC & complex coacervates & MATILDA.FT simulation package available on GitHub & \cite{Jedlinska2024EffectsCoacervates} \\
        AFINES$^{\ddagger}$ & kMC & cytoskeletal assembly & C++ package available on authors' website & \cite{Freedman2017ANetworks} \\ 
        Cytosim$^{\ddagger}$ & kMC & cytoskeletal assembly & C++ package available on GitLab & \cite{Lugo2023ASystems} \\
        aLENS$^{\ddagger}$ & kMC & cytoskeletal assembly & available on GitHub & \cite{Yan2022TowardAssemblies} \\ 
        MEDYAN$^{\ddagger}$ & kMC & cytoskeletal assembly & available on MEDYAN website & \cite{Popov2016MEDYAN:Networks} \\
    \end{tabular}
    \label{tab:table}
    \footnotesize{${\dagger}$ associative bonding mechanism, ${\ddagger}$ dissociative bonding mechanism}
\end{table}

Looking forward, the simulation of dynamic bonding in soft matter is expected to benefit from future algorithmic advancements, driven by ongoing research in several key areas. These include the development of new machine learning force fields to study covalent adaptive networks at length and time scales inaccessible to ab initio and QM/MM simulations~\cite{sun2023modeling}. Another promising direction is the discovery of machine-learning approaches that, when combined with molecular simulation, can reveal hidden relationships, yielding mechanistic insights that are otherwise inaccessible through experiments alone. A recent example demonstrates how the spatial distribution of crosslinkers in microgels can be inferred by this approach from experimentally measurable polymer density profiles, enabling the development of new predictive frameworks for swelling behavior~\cite{marin2025predicting}. Progress along these lines will be matched by efforts in the design of dynamic bonding materials with targeted properties using inverse methods~\cite{bedolla2020machine,kadulkar2022machine}, with recent successes highlighting the roles of machine learning-based virtual screening approaches~\cite{yan2023overcoming}, especially those that leverage MD simulation data~\cite{zheng2025ai,Zheng2025AcceleratedNetworks}, to direct the discovery process.     

\section*{Declaration of Competing Interest}
\label{competing}
The authors declare that they have no known competing financial interests or personal relationships that could have appeared to influence the work reported in this paper.

\section*{Acknowledgment}
\label{acknowledgment}
This work was supported by the National Science Foundation through a DMREF grant (CBET-2323482) and through the Center for Dynamics and Control of Materials: an NSF Materials Research Science and Engineering Center (NSF MRSEC) under Cooperative Agreement DMR-2308817.

%\bibliography{references}

\end{document}